\documentclass[twocolumn]{aastex631}

\usepackage{units}
\usepackage{amsmath}
\usepackage[mathlines]{lineno}

\shorttitle{\textit{R}-process rain}
\shortauthors{Amend et al.}

\graphicspath{{./}{figures/}}

\begin{document}

\title{\textit{R}-process Rain from Binary Neutron Star Mergers in the Galactic Halo}

\author[0000-0002-9930-3591]{Benjamin Amend}
\affiliation{Department of Physics and Astronomy \\
Clemson University \\
Clemson, SC 29634-09781, USA}

\author[0000-0002-1895-6516]{Jonathan Zrake}
\affiliation{Department of Physics and Astronomy \\
Clemson University \\
Clemson, SC 29634-09781, USA}

\author[0000-0002-8028-0991]{Dieter H. Hartmann}
\affiliation{Department of Physics and Astronomy \\
Clemson University \\
Clemson, SC 29634-09781, USA}

\begin{abstract}
Compact binary mergers involving at least one neutron star are promising sites for the synthesis of \textit{r}-process elements found in stars and planets. However, mergers can take place at significant offsets from their host galaxies, with many occurring several kpc from star-forming regions. It is thus important to understand the physical mechanisms involved in transporting enriched material from merger sites in the galactic halo to the star-forming disk. We investigate these processes, starting from an explosive injection event and its interaction with the halo medium. We show that the total outflow mass in compact binary mergers is too low for the material to travel to the disk in a ballistic fashion. Instead, the enriched ejecta is swept into a shell, which decelerates over $\unit[\lesssim 10]{pc}$ scales and becomes corrugated by the Rayleigh-Taylor instability. The corrugated shell is denser than the ambient medium, and breaks into clouds which sink toward the disk. These sinking clouds lose thermal energy through radiative cooling, and are also ablated by shearing instabilities. We present a dynamical heuristic that models these effects to predict the delay times for delivery to the disk. However, we find that turbulent mass ablation is extremely efficient, and leads to the total fragmentation of sinking \textit{r}-process clouds over $\unit[10 - 100]{pc}$ scales. We thus predict that enriched material from halo injection events quickly assimilates into the gas medium of the halo, and that enriched mass flow to the disk could only be accomplished through turbulent diffusion or large-scale inflowing mass currents.
\end{abstract}



\section{Introduction} \label{sec:intro}

Approximately half of the chemical species heavier than iron are synthesized via the rapid neutron capture process, or \textit{r}-process, in which seed nuclei are driven to the neutron drip line by successive capture events that occur before $\beta$-decay can take place \citep{burbidge_synthesis_1957, cameron_heavy_1982, seeger_nucleosynthesis_1965}. Astrophysical sites which possess the extremely high neutron fluxes, mass densities, and temperatures needed to sustain \textit{r}-process nucleosynthesis include core-collapse supernovae \citep[CCSNe;][]{nishimura_r_2015, siegel_collapsars_2019, yong_r-process_2021} and compact binary mergers involving at least one neutron star \citep[]{freiburghaus_[clc][ital]r[/ital][/clc]-process_1999, goriely_r_2011, 10.1111/j.1365-2966.2012.21859.x, surman_r-process_2008}. It is a major goal of nuclear astrophysics to understand the relative contributions by these events to cosmic \textit{r}-process enrichment.

Binary neutron star (BNS) mergers are promising as major \textit{r}-process nucleosynthesis sites due to their extremely neutron-rich environments \citep[e.g.][]{wanajo_production_2014}. Injection time delays inferred from \textit{r}-process abundances in stars are also consistent with the expected gravitational wave inspiral times $\sim \unit[0.01 - 1]{Gyr}$ \citep{anand_merger_2018, mennekens_delay_2016} following the formation of double NS systems. The most direct evidence comes from the kilonova event AT2017gfo, associated with GW170817, which was confidently identified as a BNS merger \citep[]{abbott_gravitational_2017}. Modeling of the kilonova infrared through ultraviolet emission indicates \textit{r}-process nucleosynthesis was taking place in the outflowing, neutron-rich gas \citep{drout_light_2017, pian_spectroscopic_2017}. Indirect evidence for individual enrichment events likely due to BNS mergers has also been found in halo field stars \citep{ji_lanthanide_2019} and dwarf galaxies in the Milky Way halo \citep{ji_r-process_2016, naidu_evidence_2022, safarzadeh_r-process_2019}. Furthermore, the inferred abundances of several radioactive isotopes from the early solar system cannot be explained solely by collapsars and other types of supernovae, but may be more reflective of the yields associated with BNS mergers \citep[]{bartos_nearby_2019, hotokezaka_short-lived_2015}. Population synthesis studies indicate that the estimated GW event density rates are consistent with galactic BNS merger rates needed to explain observed \textit{r}-process abundances \citep[]{cote_origin_2018}.

Many BNS mergers occur at substantial offsets from their host galaxies, potentially occurring at high galactic latitudes well separated from any star-forming regions. Evidence for this comes mainly from short-duration gamma-ray bursts (sGRBs), which are produced by jetted relativistic outflows in BNS mergers \citep{abbott_gravitational_2017, doi:10.1146/annurev-astro-081913-035926}. The burst locations are typically found to have projected offsets in the range of $\unit[1-10]{kpc}$ from the host galaxy, when one is identified \citep{bloom_observed_2002, fong_hubble_2010, Fong2013}. In particular, results from \citet{Fong2013} indicate that BNS mergers may take place almost exclusively in regions far separated from the star forming ISM.
It is thus important to understand the physical mechanisms of mass transport between the putative \textit{r}-process injection sites in galactic halos, and any star-forming regions in the host galaxy where heavy elements can be eventually incorporated into new generations of stars.

In this study, we aim to elucidate key physical processes that characterize the delivery of nuclear species produced by BNS mergers in the halo to active star forming regions in the disk. We start by discussing the dynamics of explosive enriched outflows interacting with the halo medium. We argue that the explosion material is initially confined to compact regions of the halo, but subsequently forms overdense clouds which sink toward the disk. The sinking enriched clouds are affected by drag and buoyancy forces, in addition to radiative cooling and turbulent mass ablation, and we develop model dynamical equations to account for these processes. We then explore numerical solutions to these equations, to assess the feasibility of delivering \textit{r}-process enriched gas clouds to the ISM from high-altitude injection sites.

Our paper is organized as follows. Sec. \ref{sec:explosions} enumerates the possible modes by which nucleosynthesis products could be transferred from halo injection sites to the galactic disk. In Sec. \ref{sec:model}, we present our dynamical model for sinking enriched clouds. In Sec. \ref{sec:results}, we explore numerical solutions to the model, and present results for a fiducial injection scenario, motivated by parameters estimated for the BNS merger GW170817. We also explore the sensitivity of our results to variations in initial conditions. In Sec. \ref{sec:implications}, we comment on the implications of our results for contemporary GCE models. Sec. \ref{sec:summary} summarizes our findings and poses questions that will require more detailed hydrodynamic or thermo-chemical modeling to resolve in full.

\section{\textit{R}-process injection in the halo}
\label{sec:explosions}

The projected separations of short GRBs from their host galaxies indicate that many binary neutron stars merge in halo gas environments, well-separated from star-forming regions in the gas disk. The observed offsets of $\unit[0.1 - 10]{kpc}$ are consistent with kicks on the order of \unit[100]{km/s} imparted by the SN explosions of the component stars (\citealt{belczynski_effect_1999, vigna-gomez_formation_2018}; see however \citealt{perets_no_2021}), and the aforementioned long inspiral times. Various modes of mass transfer could be involved in the delivery of \textit{r}-process-enriched material to star-forming regions. These modes are summarized here.

\subsection{Diffusive Transport}
Any small-scale turbulent motions of the halo gas would cause diffusive transport of chemical species
from regions of high concentration to regions of low concentration. If the injection site is located at a distance $z_{\rm ej}$ above the disk, diffusive transport would bring a good fraction of the material to the disk on a timescale $\tau_{\rm diff} \sim z_{\rm ej}^2 / D$, where $D \equiv v_{\rm RMS} \ell_{\rm eddy}$ is the turbulent diffusion coefficient. For an RMS turbulence velocity $v_{\rm RMS} \sim \unit[10]{km \cdot s^{-1}}$ and outer scale $\ell_{\rm eddy} \sim \unit[100]{pc}$ \citep[e.g.][]{Elmegreen2004}, $D \sim \unit[1]{kpc^2 \cdot Gyr^{-1}}$. Smaller diffusion coefficients $\sim \unit[0.1]{kpc^2 \cdot Gyr^{-1}}$ have been suggested in other works such as \cite{beniamini_turbulent_2020}. Taken together with a fiducial $z_{\rm ej} \sim \unit[]{kpc}$ we estimate that $\tau_{\rm diff}$ is in the range of $\unit[1-10]{Gyr}$, but this depends on very uncertain characteristics of turbulence in the halo medium \citep{putman_gaseous_2012}.

\subsection{Advective Transport}
%
Enriched material can also be carried over large distances by the prevailing mass flow in and around a galaxy, which we refer to here as ``advective'' transport. For example, the injection of enriched material into a starburst-driven wind could result in the material being carried away to the intergalactic medium (IGM), potentially removing the possibility for the enriched material to enter the star-forming ISM \citep[]{heckman_nature_1990, veilleux_galactic_2005}. On the other hand, enriched material that is entrained in galactic inflows such as hot-mode accretion \citep[]{keres_how_2005} would be expected to join the disk over a timescale $\sim M_{\rm halo} / \textrm{SFR}$. For a fiducial halo mass $M_{\rm halo} = \unit[10^9]{M_{\odot}}$ and SFR $\sim \unit[1]{M_{\odot} \cdot yr^{-1}}$, this mode would also be also quite slow, operating over $\sim \unit[1]{Gyr}$.

\subsection{Ballistic Transport}
Since injection events are explosive, it is worth considering whether ejecta from a BNS merger can potentially travel ballistically from a halo injection site to the disk. If so, then ballistic transport would yield very short delay times on the order of $z_{\rm ej} / v_{\rm ej} \sim  \unit[10^4]{yr}$ for outflow velocities $v_{\rm ej} \sim 0.1c$ and $z_{\rm ej} \sim \unit[]{kpc}$. However, as this material is thrown outward, it is decelerated against the halo gas over some length scale $\ell_{\rm dec}$; as such, this mode of transport is only feasible if $z_{\rm ej} \ll \ell_{\rm dec}$.

To determine whether ballistic mass transfer is possible, we consider an explosion which occurs in the halo, launching an outflow of mass $M_{\rm ej}$ radially. In the ballistic (or free-expansion) phase, this mass coasts outwards in a thin shell. The shell decelerates when it is crossed by the reverse shock which forms due to interaction with the ambient medium.
Deceleration occurs when the full kinetic energy of ejecta, $\sim M_{\rm ej} v_{\rm ej}^2$, is transferred to the external medium. The shocked ejecta moves slower than the unshocked ejecta by a modest factor, so its kinetic energy is on the order of $\sim M_{\rm swept} v_{\rm ej}^2$. The ejecta energy is fully transferred to the ambient medium when $M_{\rm ej} \sim M_{\rm swept}$, at which time the shell has advanced to a radius of order
\begin{linenomath*}
    \begin{eqnarray}
        \ell_{\rm dec} &\equiv& \left[ \frac{3 M_{\rm ej}}{4\pi\rho_{\rm env}(z_{\rm ej})} \right]^{1/3}\\
        &\sim& \unit[10]{pc} \times \left( \frac{M_{\rm ej}}{\unit[0.01]{M_{\odot}}} \right)^{1/3} \times \left[ \frac{\rho_{\rm env}(z_{\rm ej}) / m_p}{\unit[10^{-4}]{cm^{-3}}} \right]^{-1/3} . \nonumber
        \label{eq:ldec}
    \end{eqnarray}
\end{linenomath*}
Since merger altitudes $z_{\rm ej}$ are $\sim \unit[0.1 - 10]{kpc}$, $\ell_{\rm dec} \ll z_{\rm ej}$ and we conclude that ballistic transfer of such small amounts of gas over large $\sim \unit[]{kpc}$ scale distances is unrealistic.

\subsection{Sinking Clouds or ``Percolation''}
\label{sec:percolation}
Once the material flowing away from the explosion site has decelerated, it may still sink through the halo. This is because the Rayleigh-Taylor (RT) instability leads to corrugation of the contact discontinuity separating the shocked merger ejecta from the shocked ambient medium. Because the shocked ejecta is colder and denser than the shocked ambient medium, the plumes are over-dense, and begin to sink vertically downwards from the explosion site. We refer to the sinking enriched plumes as ``clouds'', and use the symbol $\ell$ to characterize their instantaneous linear dimension.

When they first form, these clouds have the linear dimension of the RT plumes associated with the explosion $\ell \sim \ell_{\rm dec} \theta_{\rm RT}$, where $\theta_{\rm RT} \ll 1$ is the angular scale of the fastest growing RT unstable modes. The initial density $\rho_{\rm ej,0}$ of the the shocked gas in the RT plumes can be approximated as the total ejecta mass divided by the volume of the RT-corrugated shell, $\rho_{\rm ej,0} = M_{\rm ej} / V_{\rm shell}$. The RT instability saturates when the shell is at radius $\ell_{\rm dec}$ and has volume $V_{\rm shell} = 4 \pi \ell_{\rm dec}^2 \delta r$, where the shell width is comparable to the linear scale of the plumes, $\delta r \sim \theta_{\rm RT} \ell_{\rm dec}$. Using Eq. \ref{eq:ldec}, the initial cloud density is seen to be higher than that of the ambient medium,
\begin{equation}
    \rho_{\rm ej,0} = \frac{\rho_{\rm env}(z_{\rm ej})}{3 \theta_{\rm RT}} \, .
    \label{eq:rho-ej0}
\end{equation}
%
%
Eq. \ref{eq:rho-ej0} is accurate when the ejecta decelerates on a length scale much smaller than the density scale height of the halo atmosphere. The mean density of the enriched clouds when they are first formed is thus higher than the ambient density by a factor inversely proportional to the RT angular scale.
Since the clouds are denser than the ambient medium, they begin to sink in the gravity of the host galaxy. We refer to this process interchangeably as sinking, or ``percolation'', in analogy with ground water seeping through porous soil.

The dynamics of sinking enriched clouds are affected by drag, buoyancy, radiative cooling, and mass loss associated with shear instabilities. These processes are non-trivial in conjunction, and no simple estimate of the associated timescale can be made. Indeed, there is also the possibility that shear instabilities could lead to the total assimilation of the cloud material into the ambient medium --- an outcome that would make percolation ineffective at transporting enriched material to any star-forming ISM at low altitudes. In order to assess how effective the sinking clouds might be at transferring enriched material over large distances, we propose in the next section a toy model to predict their evolution as they sink in a stratified gas medium.


\section{Dynamics of a Sinking Cloud}
\label{sec:model}

\subsection{The Galactic Environment}
\label{sec:galenv}

We model the galactic environment as a horizontally-stratified medium composed of atomic hydrogen, which transitions smoothly between ISM and halo number densities. The midplane of the disk has number density $n_{\rm disk}$ and nominal pressure scale height $b$. The mass density varies with altitude as
\begin{linenomath}
    \begin{equation*}
        \rho_{\rm env}(z) = n_{\rm disk} m_p \left( 1 + \frac{z^2}{b^2} \right)^{-1} \, .
        \label{eq:atmospheredensity}
    \end{equation*}
\end{linenomath}
This vertical density profile is smooth at the galactic midplane, and captures measured values of the ISM and halo mass density with reasonable accuracy. It is also matches extremely closely the functional form chosen by \cite{ferriere_hot_1998} for the purpose of modeling the expansion of supernova remnant hot bubbles through the upper layers of the ISM. The pressure profile
\begin{linenomath}
    \begin{equation*}
        P_{\rm env}(z) = n_{\rm disk} m_p bg \left[ \frac{\pi}{2} + \tan^{-1}{\left( \frac{\pi}{2} - \frac{z}{b} \right)} \right]
        \label{eq:atmospherepressure}
    \end{equation*}
\end{linenomath}
follows from the hydrostatic equilibrium condition,
\begin{linenomath}
\begin{equation*}
    \frac{dP_{\rm env}}{dz} = -\rho_{\rm env} g(z) \, .
\end{equation*}
\end{linenomath}
Following \cite{benjamin_high-velocity_1997} and \cite{wolfire_multiphase_1995}, we approximate $g(z) = {\rm const} = \unit[10^{-8}]{cm \cdot s^{-2}}$.

\begin{figure}
	\includegraphics{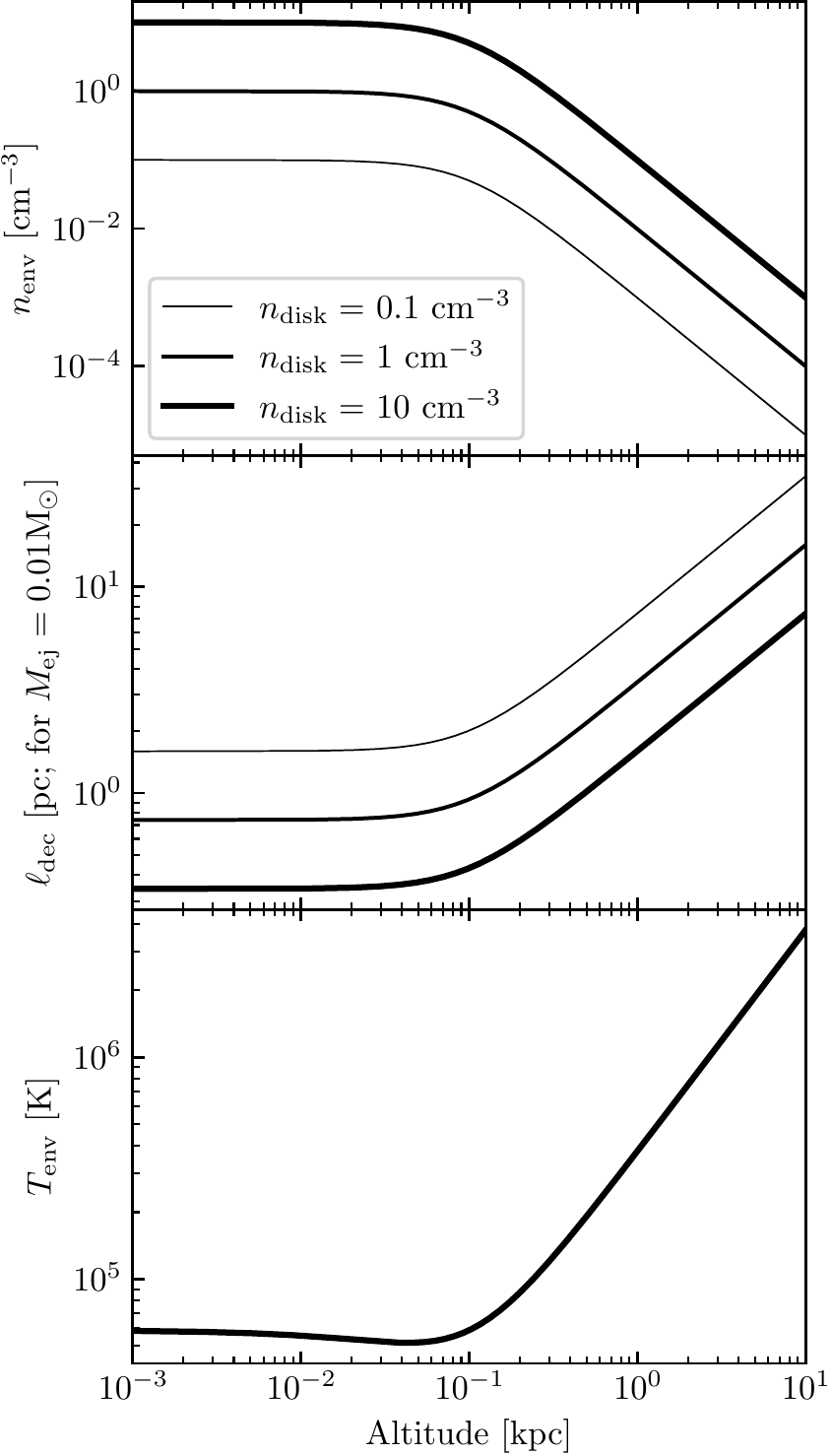}
    \caption{Vertical profile of the hydrostatic equilibrium profile we adopted for the halo atmosphere, for representative values of the number density $n_{\rm disk}$ at the base of the atmosphere. The top panel shows the number density $n_{\rm env}$ as a function of altitide. The middle panel shows the corresponding nominal deceleration length scale $\ell_{\rm dec}$ (Eq. \ref{eq:ldec}) for the fiducial outflow mass $M_{\rm ej} = 0.01 M_\odot$. The bottom panel shows the vertical temperature profile. Note that it is insensitive to the density normalization.}
    \label{fig:atmosphere}
\end{figure}

The density profile and corresponding temperature profiles are plotted in Fig. \ref{fig:atmosphere} for three different values of $n_{\rm disk}$. For reference, Fig. \ref{fig:atmosphere} also shows the nominal deceleration length $\ell_{\rm dec}$ from Eq. \ref{eq:ldec} as a function of altitude, for the fiducial outflow mass $M_{\rm ej} = \unit[10^{-2}]{M_{\odot}}$. Note that for explosion altitudes in the range $z_{\rm ej} = \unit[0.1 - 10]{kpc}$, the nominal deceleration length scale is much smaller than the density scale height $b = \unit[100]{pc}$, so the vertical density gradient of the atmosphere does not significantly influence the geometry of the decelerating ejecta shell. Also note that the ambient temperature is relatively uniform at low altitudes $z \lesssim b$, $k_B T_{\rm env} \simeq m_p g b$.

\subsection{Cloud Composition}
\label{sec:composition}
Sinking clouds are composed of a mixture of enriched BNS merger ejecta and ambient halo gas, which is mostly hydrogen. The mixture is formed due to turbulent mixing across the contact discontinuity brought about by the RT instability of the decelerating shell. The composition of the BNS merger outflow includes some $r$-process heavy elements, but the majority of the ejecta is composed of lighter elements. The composition of the mixture of the enriched ejecta and halo gas is characterized by the effective (mean) atomic weight,
\begin{equation}
    A \equiv \frac{\Sigma_i n_i A_i}{\Sigma_i n_i} \, ,
    \label{eqn:effective-A}
\end{equation}
where the sum is over the nuclear species and $n_i$ and $A_i$ are the number density and atomic mass of species $i$. The nucleosynthetic processes determining the outflow composition are influenced by the expansion history \citep{Qian1996, Kuroda2008}, and also by shock-heating and turbulent mixing in the cocoon of a relativistic jet if one is produced \citep{Hotokezaka2013,Gottlieb2018,Xie2018}.
Given the large uncertainty in the nucleosynthetic yields, and also how much RT mixing dilutes the enriched ejecta with ambient hydrogen, we consider the effective atomic mass $A$ of the clouds to be very uncertain, and generate sample results for values in the range of $A = 2 - 160$. Multi-dimensional hydrodynamic simulations of passive scalar mixing across an RT-unstable contact discontinuity will be needed to determine the dilution factor.



We will assume in our analysis that enriched clouds are composed mainly of ionized gas. As we will see in Sec. \ref{sec:results} the temperature of the sinking gas clouds is typically in the range of $\unit[10^{5-6}]{K}$, above the threshold for hydrogen ionization. These high temperatures are the result of the enriched clouds being in pressure equilibrium with the halo atmosphere, and are higher than in the kilonova free-expansion phase (the ejecta temperature increases abruptly when it is crossed by the reverse shock). At such high temperatures, we do not expect any dust grains, which might have formed in the nebular kilonova phase \cite[e.g.][]{takami_dust_2014}, to have survived, however this expectation should be tested with non-equilibrium thermo-chemical models.


\subsection{Forces on Sinking Clouds}
\label{sec:dragbuoyancy}

Sinking gas parcels are subject to the gravitational force, the buoyancy force, and the drag force,
\begin{equation}
    F_{\rm net} = -m g + \rho_{\rm env} V g - v^2 \rho_{\rm env} a \times \textrm{sgn}(v) \, .
    \label{eq:cloudnetforce}
\end{equation}
Here, $m$ is the cloud mass, $V = 4\pi\ell^3 / 3$ is the cloud volume, $\ell$ is the cloud radius, $a = \pi \ell^2$ is the cross-sectional area, and $v = \dot z$ is the cloud velocity. For sufficiently overdense sinking clouds, the acceleration is dominated by the gravitational force, and the downward trajectory is essentially free-fall. In that limit, the timescale for the cloud to fall from altitude $z$ to the disk would be
\begin{equation}
    \tau_{\rm fall} = \sqrt{\frac{2z}{g}} \, .
    \label{eq:freefalltime}
\end{equation}
In general, the drag force has an associated stopping timescale
\begin{equation}
    \tau_{\rm stop} \equiv \frac{m |v|}{v^2 \rho_{\rm env} a} \, .
    \label{eq:terminalvelocitytime}
\end{equation}
When $\tau_{\rm stop} \ll \tau_{\rm fall}$, clouds would spend most of the descent moving at the terminal velocity,
\begin{equation}
    v_{\rm term} = \sqrt{\frac{4g\ell}{3} \left( \frac{\rho}{\rho_{\rm env}} - 1 \right)} \, ,
    \label{eq:terminalvelocity}
\end{equation}
where $\rho = m / V$. Eq. \ref{eq:terminalvelocity} is obtained by setting the left-hand-side of Eq. \ref{eq:cloudnetforce} to zero.

Eq. \ref{eq:cloudnetforce} has exact solutions that act like a damped oscillator for small oscillations about the neutral buoyancy altitude. However, in a realistic setting, clouds experience mass loss due to ablation, and energy loss due to radiative cooling of the ionized plasma. These effects are important to the overall mass transfer process, and we account for them in the following subsections.

Note that we have treated the clouds as sinking through a medium characterized by the assumed halo profiles $\rho_{\rm env}(z)$ and $P_{\rm env}(z)$. This is only accurate if the kilonova shock wave did not permanently or significantly alter those profiles. The shock runs ahead of (i.e. is at lower altitude than) the contact discontinuity, where the enriched clouds are formed from the RT plumes. Because the shock moves so much faster than the contact, we expect that the halo medium will have recovered to its ambient conditions, near hydrostatic equilibrium, by the time the clouds sink downwards. This picture should be checked with detailed hydrodynamics simulations of explosions in stratified media.

\subsection{Mass Ablation by Shear Instabilities}
\label{sec:fragmentation}

As the cloud falls it is subject to Kelvin-Helmholtz (KH) instabilities associated with the velocity shear between the sinking material and the halo gas. These instabilities remove mass from the cloud and disperse it into the medium through which it is sinking. We refer to this process as ablation, and characterize its effect as removing mass at the KH instability growth rate,
\begin{equation}
    \gamma_{\rm KH} = k v \frac{\sqrt{\rho \rho_{\rm env}}}{\rho + \rho_{\rm env}} \, ,
    \label{eq:khrate}
\end{equation}
where $k$ is the spatial frequency of the growing KH mode. Although the small-scale perturbations grow faster, it is the perturbation with spatial frequency $k \sim \ell^{-1}$ which will ultimately break the cloud apart. The timescale associated with this breakup is then
\begin{equation}
    \tau_{\rm KH} = \frac{\ell}{|v|}\frac{\rho + \rho_{\rm env}}{\sqrt{\rho\rho_{\rm env}}} \, ,
    \label{eq:khtime}
\end{equation}
where $\ell$ corresponds to the (time-evolving) linear dimension of the RT structures described in \ref{sec:percolation}. The evolution of the cloud mass accounting for turbulent ablation is then governed by
\begin{linenomath*}
\begin{equation}
    \dot m(t) = - m(t) \times \tau_{\rm KH}^{-1} \, .
    \label{eq:dmblobdt}
\end{equation}
\end{linenomath*}

\subsection{Radiative Cooling}
\label{sec:cooling}

For our fiducial cloud composition, plasma temperatures are $\sim \unit[10^5 - 10^6]{K}$, as we will show in Sec. \ref{sec:heuristic}; as such, the cloud material is highly ionized. The clouds are optically thin, with electron-scattering optical depth $\tau_{\rm es} = n_e \sigma_T \ell \sim 10^{-7}$. We thus adopt free-free emission as a first approximation to the cloud radiative cooling. Some caveats are noted below. The frequency-dependent emission (in $\unit[]{erg \cdot cm^{-3} \cdot s^{-1} \cdot Hz^{-1}}$) due to thermal free-free radiation of a plasma with ion density $n_i$ and electron density $n_e$ is
\begin{linenomath}
\begin{equation*}
    \dot{u}_{\nu} = \frac{2^5 \pi e^6}{3m_ec^3}\left( \frac{2\pi}{3k_Bm_e} \right)^{1/2}T^{-1/2}Z^2n_en_ie^{-h\nu / k_BT}\bar{g}
    \label{eq:freefreeemission}
\end{equation*}
\end{linenomath}
\citep[]{rybicki_lightman_2004}, where $Z$ is the effective atomic number of the cloud plasma. This can be integrated over all frequencies to obtain the radiated power per unit volume ($\unit[]{erg \cdot cm^{-3} \cdot s^{-1}}$),
\begin{linenomath}
\begin{equation*}
    \dot{u} = \frac{2^5\pi e^6}{3hm_ec^3}\left( \frac{2\pi k_B}{3m_e} \right)^{1/2}T^{1/2}Z^2n_en_i\bar{g} \, .
\end{equation*}
\end{linenomath}
The gaunt factor $\bar{g}$ is of order unity, so it will be omitted from this point forward. The ion number density is
\begin{linenomath}
\begin{equation*}
    n_i = \frac{\rho}{m_pA} \, ,
\end{equation*}
\end{linenomath}
where $A$ is the effective atomic mass number. In terms of the ionization fraction $\alpha$, the electron number density is
\begin{linenomath}
\begin{equation*}
    n_e = \alpha n_i Z.
\end{equation*}
\end{linenomath}
In order to account for the suppression of free-free cooling below the hydrogen recombination temperature $T_{\rm min} \equiv \unit[10^4]{K}$, we model $\alpha$ as a step function, $\alpha(T|T_{\rm min}) = \Theta(T - T_{\rm min})$. In obtaining numerical solutions, we use a small amount of smoothing to improve the robustness of the Runge-Kutta integrations. The radiated power per volume $\dot{u}$ can now be expressed as
\begin{equation}
    \dot{u} = \frac{2^5 \pi e^6}{3hm_em_p^2c^3}\left( \frac{2k_B\pi}{3m_e} \right)^{1/2}\frac{\alpha(T|T_{\rm min}) Z^3 T^{1/2}\rho^2}{A^2},
    \label{eq:freefree}
\end{equation}
and the energy density $u$ is
\begin{linenomath}
\begin{equation*}
    u = \frac{3}{2}(n_e + n_i)kT \, .
\end{equation*}
\end{linenomath}
Clouds lose a significant fraction of their thermal energy over the cooling timescale, $\tau_{\rm cool} \equiv u/\dot{u}$. In Sec. \ref{sec:results} we examine the relative importance of radiative cooling and mass ablation by comparing the time scales $\tau_{\rm cool}$ and $\tau_{\rm KH}$ (Eq. \ref{eq:khtime}).

At temperatures around $\unit[10^5]{K}$ it is likely that the heavy elements will not be fully ionized, and could undergo faster cooling by line emission. Therefore if the gas is composed substantially of heavier elements, the free-free cooling prescription could underestimate the radiative cooling rate. To explore the dynamical effects of very fast radiative cooling,
we use the cloud composition $A$ as a cooling amplification factor, noting that the specific radiated power $\dot{u}$ scales as $Z^3/A^2 \sim Z \sim A$ (Eq. \ref{eq:freefree}). High values of $A$ can more readily trigger catastrophe cooling, as discussed in Sec. \ref{sec:results}.

\subsection{A Heuristic for the Dynamics of Sinking Clouds}
\label{sec:heuristic}
Here we present our toy model of the dynamical evolution of sinking $r$-process enriched clouds injected by high-altitude BNS merger events. The state of a cloud is characterized by four variables: the cloud altitude $z$, vertical velocity $v$, mass $m$, and energy-per-particle $\epsilon$. A single cloud of mass $m$ is representative of an ensemble of identical clouds whose masses always sum to $M_{\rm ej}$. In this sense, the process of mass ablation is equivalent to fragmentation of the cloud into smaller clouds, increasing the ensemble population to satisfy the total mass constraint.

Changes in $z$ and $v$ are described by Eq. \ref{eq:cloudnetforce}, and mass ablation takes place according to Eq. \ref{eq:khrate}. The-energy per-particle evolves according to the first law,
\begin{equation}
    d\epsilon = \delta \mathcal{Q} - Pd\mathcal{V} \, ,
    \label{eq:firstlaw}
\end{equation}
where $\mathcal{V}$ is the volume per particle. $\delta \mathcal{Q}$ accounts for radiative losses as discussed in Sec. \ref{sec:cooling}, and is intrinsically negative. $Pd\mathcal{V}$ is the pressure-volume work done by the cloud on the surroundings. It is intrinsically negative when the cloud is adiabatically heated from sinking into denser layers of the halo atmosphere.

Clouds are assumed to remain in pressure equilibrium with their surroundings as they sink, so $P = P_{\rm env}(z)$. This pressure $P$ is also related to $\epsilon$ and $n = N/V$ through the gamma-law equation of state,
\begin{linenomath}
\begin{equation}
    P = \epsilon n(\Gamma - 1) \, ,
    \label{eq:eos}
\end{equation}
\end{linenomath}
where $\Gamma = 5/3$ is the adiabatic index, and $N = m / (m_p A)$. The cloud density is $\rho = m / V$, and the temperature is defined through Eq. \ref{eq:eos} via $T = P / (n k_B)$. Together with the condition of hydrostatic equilibrium, Eq. \ref{eq:firstlaw} can be rewritten in terms of time-differentiated quantities as
\begin{linenomath}
\begin{equation*}
    \dot{\epsilon} = \frac{\dot{\mathcal{Q}}}{\Gamma} - \frac{\rho_{\rm env}gv}{n\Gamma} \, .
\end{equation*}
\end{linenomath}
$\dot{Q}$ is precisely $-\dot{u} / n$, the energy-per-particle lost due to free-free radiation. We thus obtain a closed set of first-order ordinary differential equations for all of the state variables:
\begin{align}
    \dot{z} &= v, \label{eq:heuristic1}\\
    \dot{v} &= -g\left(1 - \frac{\rho_{\rm env}}{\rho}\right) - \frac{\rho_{\rm env}}{\rho}\frac{v^2}{\ell}\times\textrm{sgn}(v), \label{eq:heuristic2}\\
    \dot{m} &= -\frac{|v|}{\ell} \frac{\sqrt{\rho \rho_{\rm env}}}{\rho + \rho_{\rm env}}m, \label{eq:heuristic3}\\
    \dot{\epsilon} &= -\frac{\dot{u}}{n\Gamma} - \frac{\rho_{\rm env} g v}{n\Gamma} \label{eq:heuristic4}\,.
\end{align}
Eqs. \ref{eq:heuristic1}-\ref{eq:heuristic4} model the dynamics of sinking \textit{r}-process-enriched clouds, including their trajectory and thermodynamic evolution, following the explosion and deceleration phases of the BNS merger event. In the next section, we present numerical solutions to this model system, and explore the outcomes to determine whether sinking clouds can be effective vehicles for the delivery of enriched material from \textit{r}-process production sites in the halo, to star-forming regions in the disk.

\section{Results}
\label{sec:results}
Numerical solutions were obtained for the dynamical heuristic developed in the previous section, Eqs. \ref{eq:heuristic1} through \ref{eq:heuristic4}. We have chosen a fiducial model motivated by inferences made for GW170817: the explosion altitude is $z_{\rm ej} = \unit[2]{kpc}$ \citep[]{abbott_progenitor_2017} and the ejecta mass is $M_{\rm ej} = \unit[10^{-2}]{M_{\odot}}$ \citep[]{abbott_estimating_2017}. The angular RT scale is selected to be $\theta_{\rm RT} = 0.1$, motivated by recent supernova remnant simulations \citep{Polin2022, Porth2014}. The effective atomic number of the outflowing material
is given the fiducial value $A = 2$, and heavy composition cases of $A = 100$ and $A = 160$ are also computed for comparison.
In this section, we present results from this fiducial model, and also explore the sensitivity of the model to different initial conditions, giving special attention to variations in the effective composition, as these influence the cooling efficiency.

\begin{figure}
	\includegraphics[width=\columnwidth]{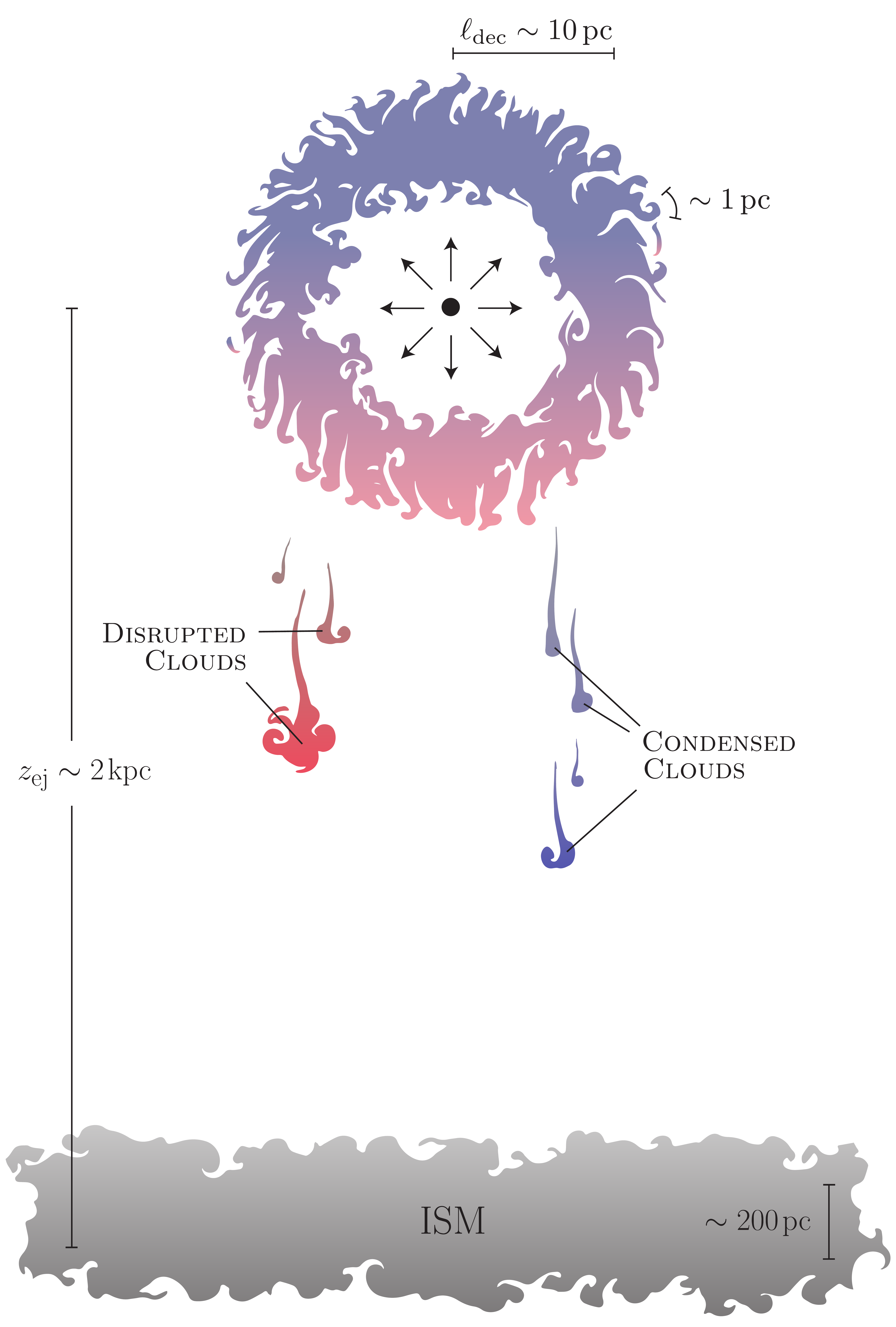}
    \caption{Different regimes of sinking \textit{r}-process enriched clouds. A BNS merger takes place at an altitude $z_{\rm ej}$ above the midplane of the galactic disk. Material expelled during the merger forms a shell, which decelerates over a length scale $\ell_{\rm dec}$ (Eq. \ref{eq:ldec}). As it decelerates, the shell becomes corrugated by the Rayleigh-Taylor (RT) instability. The RT structures are denser than the ambient medium, which causes them to begin sinking toward the disk. Clouds which cool slowly (depicted in red) are generally disrupted by Kelvin-Helmholtz (KH) instabilities before they experience significant radiative energy losses. Denser clouds (or those with higher metallicity; depicted in blue) may experience runaway free-free cooling. However once the temperature drops low enough, the plasma forms neutrals and free-free cooling is suppressed. The cold, dense, neutral clouds are still prone to fragmentation by KH. For any conditions applicable to BNS mergers at altitudes $\gtrsim \unit[100]{pc}$ above the disk, enriched material is fully dissolved into the halo gas. It can only be transported to the ISM through turbulent diffusion (if turbulence operates in the halo) or by large-scale inflowing mass currents.}
    \label{fig:graphic}
\end{figure}

\begin{figure}
    \includegraphics{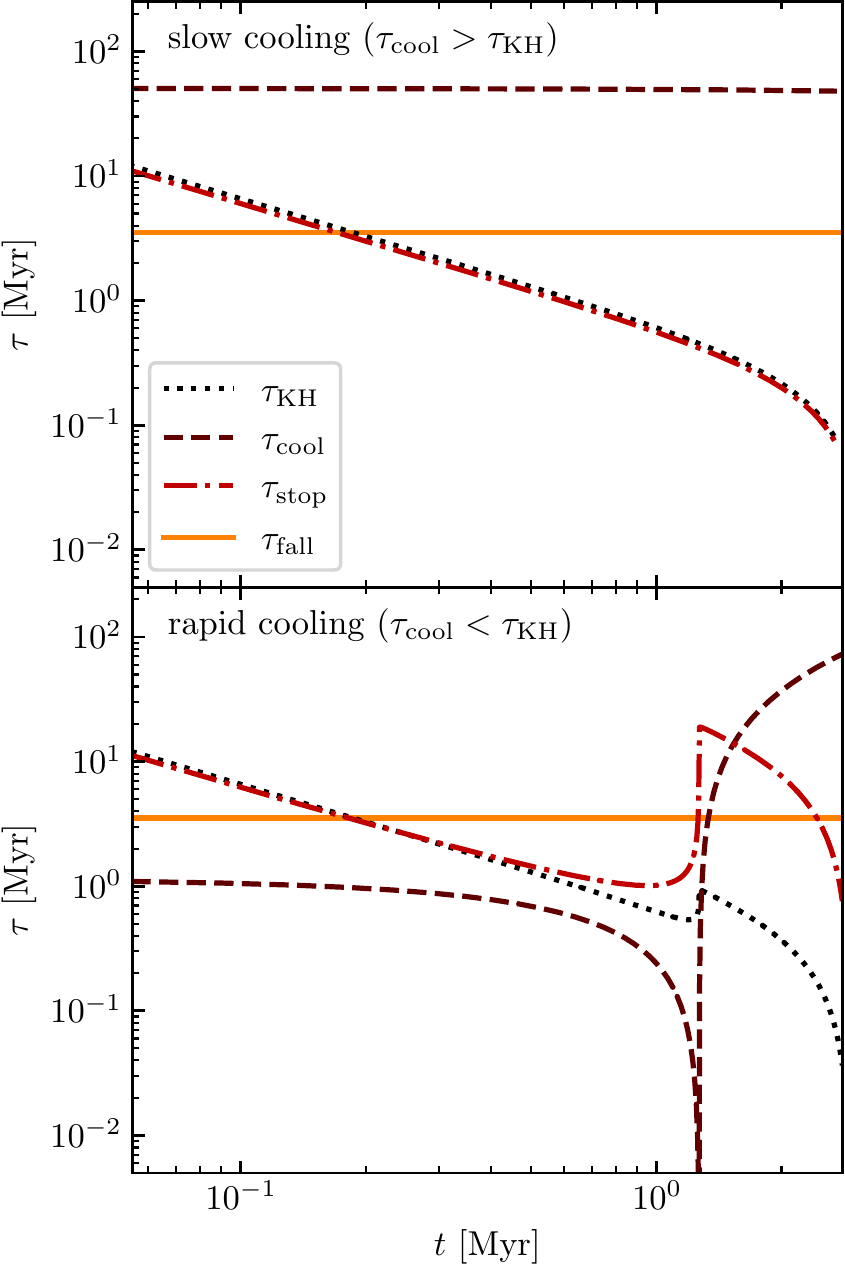}
    \caption{Four different timescales associated with solutions to the dynamical heuristic from Sec. \ref{sec:heuristic}. The upper panel shows a calculation with the fiducial injection parameters, where the clouds start out and remain in the slow-cooling regime. The lower panel uses a more metal-rich composition, $A = 100$, corresponding to faster-cooling clouds. After the clouds cool and condense, radiative cooling is suppressed by recombination. The clouds then sink more rapidly, and are ultimately disrupted by the KH instability.}
    \label{fig:timescalecomp}
\end{figure}

As discussed in Sec. \ref{sec:explosions}, the deceleration of merger ejecta leads to overdense RT-corrugated structures that begin to sink in the halo (Fig. \ref{fig:graphic}). The nominal free-fall time (Eq. \ref{eq:freefalltime}, Fig. \ref{fig:timescalecomp}) for these clouds to reach the disk from $z_{\rm ej} = \unit[2]{kpc}$ is $\tau_{\rm fall} \simeq \unit[3-4]{Myr}$. However, as the clouds fall, they experience drag from the halo gas, and approach terminal velocity. In the absence of fragmentation, clouds of linear dimension $\sim \unit[1]{pc}$ and mass $\sim \unit[10^{-4}]{M_{\odot}}$ would reach velocities of $\sim \unit[100]{km \cdot s^{-1}}$ (Eq. \ref{eq:terminalvelocity}), with associated delivery times $\sim \unit[10^7]{yr}$ (Eq. \ref{eq:terminalvelocitytime}). The velocity shear that develops as the clouds pick up speed, however, contributes to an increasing KH growth rate (Eq. \ref{eq:khrate}). As larger-wavelength KH modes grow and become non-linear, the clouds fragment into smaller clouds, which in turn have have shorter KH fragmentation times due to their smaller scales, and the process quickly terminates. For the fiducial model parameters, this leads to assimilation on relatively short timescales $\sim \unit[10^6]{yr}$, and no delivery to the disk.

\begin{figure}
    \includegraphics{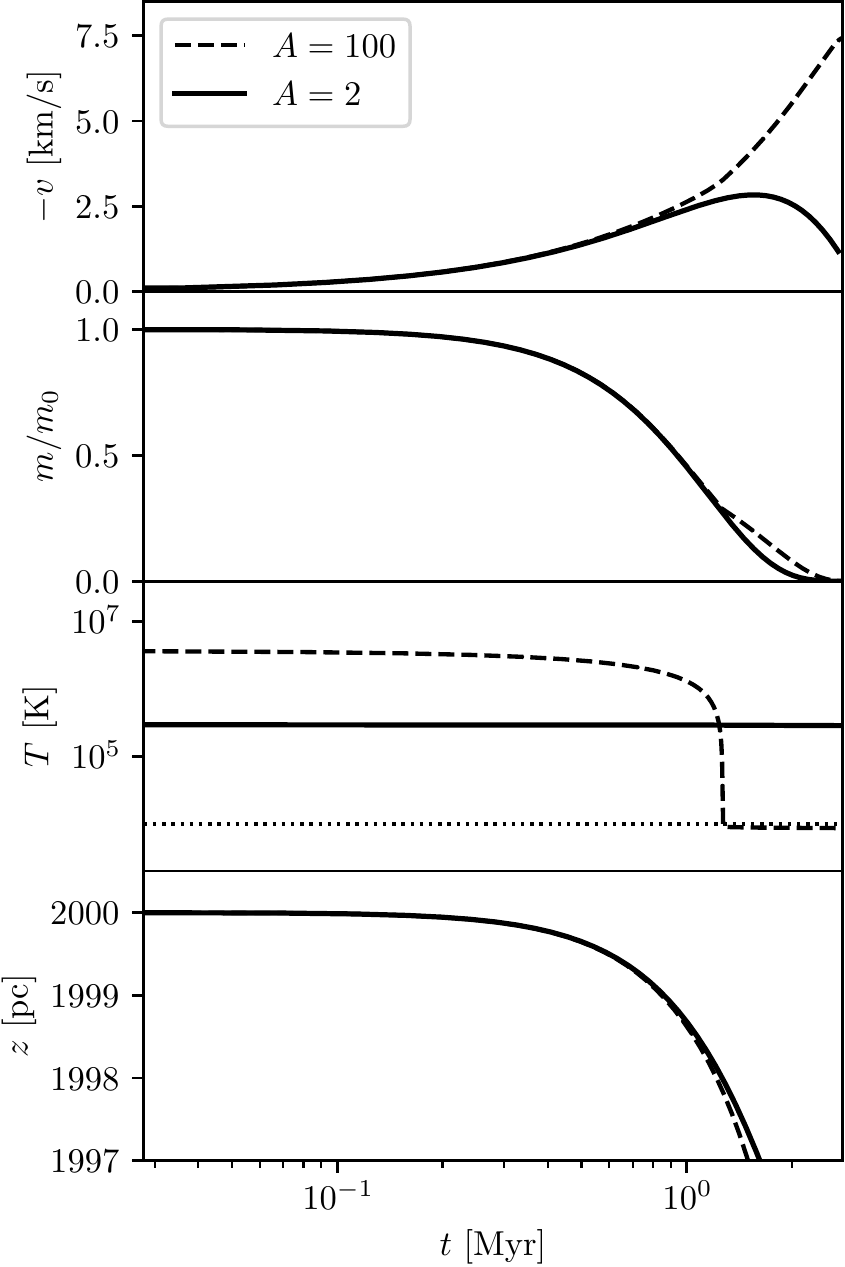}
    \caption{The evolution of the cloud velocity (top), normalized cloud mass (upper middle), temperature (lower middle), and altitude (bottom) for two different cloud compositions, corresponding to the cases presented in Fig. \ref{fig:timescalecomp}. More metal-rich material cools faster. The horizontal dashed line in the middle panel shows the cooling threshold temperature $T_{\rm min}=\unit[10^4]{K}$.}
    \label{fig:masses}
\end{figure}

For most of the cloud lifetime, the cooling timescale $\tau_{\rm cool}$ for these values is longer than $\tau_{\rm KH}$ (Fig. \ref{fig:timescalecomp}, upper panel) and as such, cooling processes are subdominant in this regime, though we explore rapid cooling solutions in detail later in this section. The complete fragmentation due to KH instabilities is seen in the numerical solutions as the cloud mass $m$ going to zero, as shown in the second panel of Fig. \ref{fig:masses}. The other panels in Fig. \ref{fig:masses} show the cloud velocity, temperature, and altitude as functions of time. Note that the temperatures shown in the third panel of Fig. \ref{fig:masses} are higher than what are predicted in early phases ($\lesssim \unit[100]{yr}$) of kilonova remnant evolution \citep[e.g.][]{Rosswog2014}. This is because enriched clouds only begin to sink after they have been decelerated and heated by the passage of a reverse shock which establishes pressure equilibrium between the clouds and the ambient halo gas. The deceleration time scale $t_{\rm dec} \equiv \ell_{\rm dec} / v_{\rm ej}$ is nominally $\sim \unit[30]{yr}$, and the cloud temperatures $\unit[10^{5-6}]{K}$ we have computed apply to times much later than $t_{\rm dec}$.

\begin{figure}
    \includegraphics{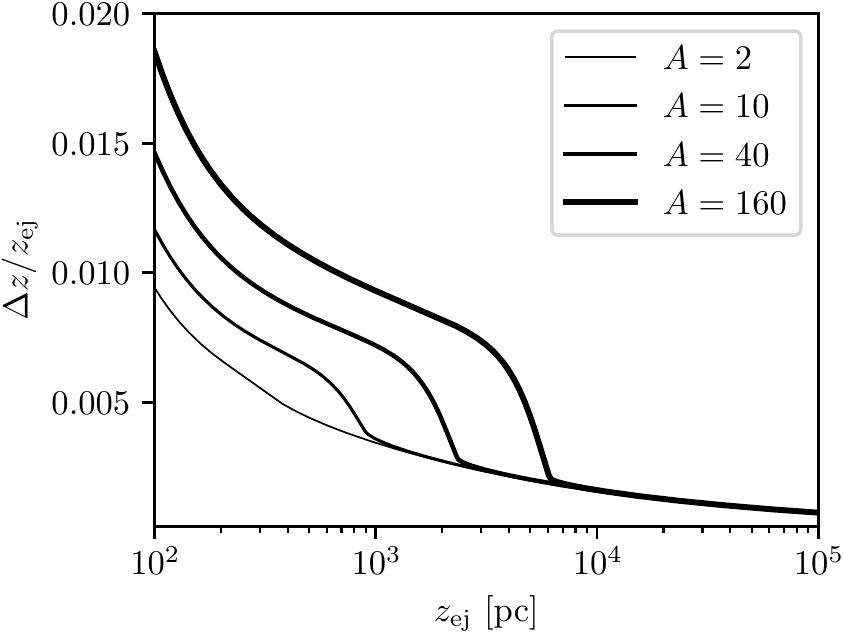}
    \caption{The fractional vertical distance below the explosion altitude $z_{\rm ej}$ at which total assimilation of the \textit{r}-process clouds takes place, as a function of $z_{\rm ej}$. Different curves represent different cloud compositions, with larger $A$ corresponding to less efficient dilution by the ambient halo medium. These calculations are for the fiducial values of $M_{\rm ej} = \unit[10^{-2}]{M_{\odot}}$, and $\theta_{\rm RT} = 0.1$.}
    \label{fig:zm}
\end{figure}

The robustness of these results with respect to variations in vertical offset and cloud composition is illustrated in Fig. \ref{fig:zm}. Importantly, we find that the cloud fragmentation goes to completion over scales $\Delta z$ that are at most $2\%$ of the injection $z_{\rm ej}$ altitude for a wide range of the model parameters $A$ and $z_{\rm ej}$. As shown in Fig \ref{fig:zm}, the ratio $\Delta z / z_{\rm ej}$ tends to decrease with greater altitude, and increase with effective atomic mass number $A$.
We have checked that the model does allow for the intact delivery to the disk of structures with much greater mass, $\gtrsim \unit[10^3]{M_{\rm \odot}}$, such as high velocity clouds \cite[HVCs; e.g.][]{benjamin_high-velocity_1997}. However, we do not find any combination of parameters for which sub-$M_\odot$ clouds, relevant to injection from BNS mergers, could survive their journeys to the disk.

We find that sinking clouds can take one of two pathways to assimilation: either fast or slow-cooling, contingent on their effective compositions. In the slow cooling mode ($\tau_{\rm cool} > \tau_{\rm KH}$), the clouds are disrupted by KH instabilities before a cooling time has elapsed (Fig. \ref{fig:timescalecomp}, upper panel), and the assimilating enriched material joins the halo gas in a warm, ionized state. In contrast, when it takes longer than a cooling time for the KH rate to exceed the cooling rate (where $\tau_{\rm cool} < \tau_{\rm KH}$), the cloud cools catastrophically, as shown in the lower panel of Fig. \ref{fig:timescalecomp}, and the assimilating material joins the halo gas in a cold, neutral state. In this fast-cooling regime, clouds rapidly cool down to the recombination temperature $T_{\rm min}$. Below $T_{\rm min}$, cooling is suppressed, and the temperature does not drop any further. The cold clouds continue to sink and are eventually fragmented by KH instabilities.

Examples of the cloud evolution in the fast and slow-cooling regimes can be seen in Fig. \ref{fig:masses}. These cases, corresponding to the same runs shown in Fig. \ref{fig:timescalecomp}, are differentiated by the their effective compositions; those with larger $A$ cool faster. The third panel of Fig. \ref{fig:masses} confirms that the temperature of the assimilating material is indeed hot for the slow-cooling case, and cold for the fast-cooling case. Complete assimilation takes place in both scenarios, as indicated by the cloud mass $m$ going to zero in the second panel, and at similar depths as seen in the fourth panel.

\section{Implications for GCE Models}
\label{sec:implications}

If \textit{r}-process-enriched material produced by halo injection sites does indeed assimilate rapidly into halo environment, then its return to the disk becomes entirely contingent on the efficacy of alternative modes of mass transfer. Accretion of gas from the halo to the disk has been found to take place in both hot and cold modes \citep[]{keres_how_2005, putman_gaseous_2012}, with cold-mode accretion dominating in older galaxies, as well as younger low-mass galaxies and in regions deep within dwarf galaxies in close proximity to the disk. This cold-mode accretion is facilitated by overdense clumps of cool gas that sink to the disk \citep[]{joung_gas_2012}, while hot-mode accretion is mediated by shock-heated gas in the outer regions of the halo, only collapsing inward upon being sufficiently cooled \citep[]{white_galaxy_1991}.

The overdense clumps in cold-mode accretion fall towards the disk as high-velocity clouds (HVCs), differing from the compact \textit{r}-process clouds discussed in this paper in composition, linear dimension, and mass \citep[]{benjamin_high-velocity_1997, kwak_simulations_2011}. These clouds are thought to absorb copious amounts of material from the surrounding halo gas as they sink \citep[]{gritton_condensation_2017, heitsch_mass_2022}, making them potential candidates for secondary means of delivery of \textit{r}-process-enriched material from the halo to the disk. Of course these clouds are also subject to hydrodynamical instabilities, and as such, many may be further disrupted along the way \citep[]{armillotta_survival_2017}. These condensation-disruption events could be repeated multiple times under the purview of a larger disk-halo mixing regime characterized by some pertinent mixing timescale. Galactic outflows may act against these accretion modes by driving material away from the disk \citep[]{martin_mapping_2005, veilleux_galactic_2005}, however. As such, even if some of this enriched material is able to reach the disk via accretion from the halo, one should expect long delay times associated with these processes.

\citet{tarumi_evidence_2021} found observational support for such delays in \textit{r}-process enrichment events in the abundance patterns of extremely metal-poor stars. In their work, a box of hydrogen gas was seeded uniformly with \textit{r}-process events according to various event rates, and this enriched material spread diffusively throughout the volume of the box. Stars were formed in this volume according to some star formation rate, and their [Ba/Mg] abundances were sampled and tracked against their metallicities. Using a standard BNS delay time distribution $\propto \Delta t^{-1}$, the authors were able to reproduce the mean [Ba/Mg] abundance pattern observed in extremely metal-poor stars, including the observed delay in enrichment. This delay was attributed to a combination of the intrinsic delay time distribution for the progenitors and the diffusion of \textit{r}-process-enriched material throughout the galaxy via turbulent mixing.

Similar diffusive and diffusive-like modes of mass transfer have been utilized in other recent GCE models \citep[]{beniamini_turbulent_2020, wanajo_neutron_2021}; however, a modified treatment wherein stars are formed at and around the midplane while \textit{r}-process injection sites occur in the halo might constitute an even more refined analysis of the data. This setup would be more reflective of the expected distributions of \textit{r}-process events and star-forming regions, and as such it would be illuminating to see by how much, if at all, the delay times extracted from the element abundances differ for such a prescription.

\section{Summary and Future Work}
\label{sec:summary}
Motivated by the fact that many BNS mergers occur well-separated from the gas disks of their host galaxies, we have developed a dynamical heuristic to study the fate of \textit{r}-process-enriched gas injected by BNS merger outflows into a galactic halo environment. For halo gas densities $n \gtrsim \unit[0.1]{cm^{-3}}$ and BNS outflow masses $M_{\rm ej} \lesssim 0.01 M_\odot$, merger outflows are decelerated over $\lesssim \unit[10]{pc}$. The decelerating shell is Rayleigh-Taylor unstable, and breaks into plumes of over-dense enriched gas, which sink through the halo gas in the gravity of the merger host galaxy. The sinking \textit{r}-process clouds are influenced by drag, buoyancy, radiative cooling, adiabatic compression, and mass ablation due to shear instabilities, and these processes are accounted for in our model.

We find that for physically relevant injection parameters, the clouds of enriched material are fragmented by KH instabilities over timescales $\sim \unit[10^6]{yr}$, indicating very compact enrichment sites within the galactic halo, generally far removed from star-forming regions in the disk. This material can assimilate into the halo in either a hot, ionized state, or a cold, neutral state depending on whether the cloud evolves in a slow or fast-cooling regime. Assimilated material could still be delivered to the disk advectively through hot-mode accretion, or by circulating mass currents such as the galactic fountain. In galaxies where little mass exchange occurs between halo gas environments and star-forming regions in the disk, or where the prevailing mass current is outwards, BNS mergers would face a new obstacle as effective \textit{r}-process enrichment sites. Alternatively, the inferred $\sim$Gyr time delays of could indicate the presence of turbulence in the halo gas, in which case the delay times would be connected to the eddy diffusivity.

A future work will present multi-dimensional hydrodynamics simulations of passive scalar mixing by turbulence across RT-unstable contact discontinuities, and of KH instability operating on overdense structures sinking in stratified media. These simulations may employ realistic radiative cooling models, and/or be coupled with non-equilibium thermo-chemical models to explore the formation of heavy element dust grains.
The chemical enrichment of galaxies is a highly coupled, multi-scale problem, and progress may require the use of large-scale cosmological simulations of evolving galaxies (\citealt{wiggins_location_2018, matteucci_modelling_2021} and references included). Our results suggest that it may be of particular importance to explore both observationally and with simulations the dynamics of turbulent mass transfer across the disk-halo interface.

\begin{acknowledgements}
We thank Christopher L. Fryer for many insightful discussions about the astrophysical environments of compact binary mergers. Additionally, we kindly acknowledge constructive assessment of the anonymous referee.
\end{acknowledgements}

\bibliography{main}{}
\bibliographystyle{aasjournal}



\end{document}